# CYP3A Mediated Ketamine Metabolism is Severely Impaired in Liver S9 Fractions from Aging Sprague Dawley Rats

Raphaël Santamaria[1], Marie-Chantal Giroux[2], Pascal Vachon[2,3] and Francis Beaudry[1]

[1] Groupe de Recherche en Pharmacologie Animal du Québec (GREPAQ), Département de Biomédecine Vétérinaire, Faculté de Médecine Vétérinaire, Université de Montréal, Saint-Hyacinthe, Québec, Canada

[2] Département de Biomédecine Vétérinaire, Faculté de Médecine Vétérinaire, Université de Montréal, Saint-Hyacinthe, Québec, Canada

[3] Centre de Recherche du CHU Sainte-Justine, Montréal, Québec, Canada

[*]Corresponding author:

Francis Beaudry, Ph.D.
Associate Professor in analytical pharmacology
Département de Biomédecine Vétérinaire
Faculté de Médecine Vétérinaire
Université de Montréal
3200 Sicotte
Saint-Hyacinthe, QC
Canada J2S 2M2

Email: francis.beaudry@umontreal.ca

Tel (514) 343-6111 ext. 8647

**Abstract**

Ketamine is widely used in veterinary medicine and in medicine. Ketamine is metabolized to its active metabolite norketamine principally by liver CYP3A. Drug metabolism alterations during aging have severe consequences particularly in anesthesiology and very few studies on older animals were conducted for ketamine. The objective of the present study is to assess the influence of aging on CYP3A metabolism of ketamine. Liver S9 fractions from 3, 6, 12 and 18 month old male Sprague Dawley rats were prepared and Michaelis-Menten parameters were determined for primary metabolic pathways. The derived maximum enzyme velocity (i.e. $V_{max}$) suggests a rapid saturation of the CYP3A enzyme active sites in liver S9 fractions of 18-month old rats. Observed $V_{max}$ for Liver S9 fractions from 3, 6 and 12 month old male Sprague Dawley rats were 2.39 (±0.23), 2.61 (±0.18), and 2.07 (±0.07) respectively compared to 0.68 (±0.02) for Liver S9 fractions from 18 month old male Sprague Dawley rats. Interestingly, we observed a 6 to 7 fold change in the derived $K_m$ when comparing Liver S9 fractions from 18 month old male Sprague Dawley rats with Liver S9 fractions from younger rats. Our results suggest that rat CYP3A enzyme undergoes conformational changes with age particularly in our geriatric group (e.g. 18 month rats) leading significant decrease in the rate of formation of norketamine. Moreover, our results strongly suggest a severe impairment of CYP3A ketamine mediated metabolism.

## Introduction

In recent years, the improvements in medicine, and life style changes, have increased life expectancy and consequently there is a significant increase of geriatric patients needing medical and pharmacological treatments. A very large proportion of marketed drugs were not thoroughly tested in pediatric and geriatric patients and this has led to numerous medical complications [Richardson *et al.*, 2014; Crentsil *et al.*, 2014; Hines, 2013]. Older patients certainly have more risks of suffering from adverse drug effects which may be associated with a decreased clearance (CL) caused by physiologic changes associated with aging [Meziere *et al.*, 2013; Woodhouse & Wynne, 1988]. Additionally, other studies show a significant increase of terminal half-life ($T_{1/2}$) of many drugs related to aging [Iirola *et al.*, 2012; Shi and Klotz, 2011] leading to important differences in pharmacokinetic profiles [Veilleux-Lemieux *et al.*, 2013]. Pharmacokinetic and drug metabolism alterations during aging have severe consequences particularly in anesthesiology [Meziere *et al.*, 2013]. Geriatric surgeries in animal and human are often avoided when possible since adverse effects could lead to serious consequences including death.

Recently, we performed an exhaustive comparison of ketamine and xylazine pharmacokinetics and pharmacodynamics using 3, 6, 12 and 18 month old male Sprague Dawley rats and observed significant differences in drug exposition (AUC), terminal half-life ($T_{1/2}$) and clearance (CL) in 12 and 18 months old rats [Giroux *et al.*, 2015]. Pharmacokinetic parameters ($T_{1/2}$ and AUC) significantly increased and drug CL significantly decreased with aging and these observations correlated with physiological results (anesthesia duration, reflexes, cardiac and respiratory frequencies and oxygen saturation). As shown in Figure 1, ketamine is metabolized primarily by the liver to an active metabolite, norketamine and it is excreted in the urine [Meyer and Fish, 2008]. In our previous study using rat liver S9 fractions and cDNA CYP3A1 and CYP3A2 expressed enzymes, we demonstrated that ketamine is a substrate of CYP3A (i.e. CYP3A1 and CYP3A2) [Santamaria *et al*., 2014]. The significant increase of $T_{1/2}$ and decrease of CL observed *in vivo* could possibly be associated with hampered CYP450 liver enzymes, which requires further investigation.

The objective of the present study is to assess the influence of aging on CYP3A metabolism of ketamine. Liver S9 fractions from 3, 6, 12 and 18 month old male Sprague Dawley rats were prepared and Michaelis-Menten parameters were determined for primary metabolic pathways. Midazolam, is primarily metabolized in the liver by the CYP3A to its pharmacologic active metabolite, α-hydroxymidazolam [Shimizu *et al*., 2007] and it is a preferred CYP3A substrate extensively used for *in vitro* and *in vivo* studies [Halama *et al*., 2013; Mooiman *et al*., 2013]. Midazolam was used to verify and validate CYP3A activity in liver S9 fractions prepared from tissues obtained from animals of different age groups. Drug and metabolite analyses were performed using a HPLC-MS/SRM method capable of quantifying norketamine and α-hydroxymidazolam using an isotopic dilution strategy in liver S9 fraction suspensions.

## Materials and Methods

### *Chemicals and Reagents*

Ketamine, $d_4$-ketamine, norketamine, $d_4$-norketamine, midazolam, $d_4$-midazolam and α-hydroxymidazolam and $d_4$- α-hydroxymidazolam were obtained in solution from Cerilliant (Round Rock, TX). Other chemicals, including acetonitrile, formic acid, methanol, sodium phosphate dibasic and sodium phosphate monobasic were purchased from Fisher Scientific (Ottawa, ON, Canada). Commercial rat liver S9 fractions and NADPH regeneration solutions were obtained from Corning Gentest (Tewksbury, MA, USA).

***Animal study***

Twenty-four specific pathogen free male Sprague Dawley rats from Charles River Canada (St-Constant, QC) were used for this study. Seven to nine weeks old rats (n=6/age group) were purchase and kept until they were respectively 3 and 6 months old. Twelve 8 month old rats were purchased and kept until they were 12 and 18 months of age (n=6/age group). All rats were housed in a standard laboratory animal environment. The rats had *ad libidum* access to food (2018 Teklad Global 18 % Protein Rodent Diet, Harlan Teklad, Bartonville, IL) and reverse osmosis water. They were single housed in ventilated cages (Green Line IVC Sealsafe Plus, Tecniplast, USA) changed once a week. Rats were housed on corn cob bedding (7097 corncob, Harlan Teklad, Bartonville, IL) and they had a high temperature polycarbonate rat retreat (Bioserv, Flemington, NJ) and one nylabone (Bioserv, Flemington, NJ) for environmental enrichment. The Institutional Animal Care and Use Committee approved the protocol prior to animal use in agreement with the guidelines of the Canadian Council on Animal Care [Canadian Council on Animal Care, 1993]. Rats were euthanized with $CO_2$, livers were rapidly remove and S9 fractions where prepared.

***Rat Liver S9 preparation and incubation***

For each age group, three livers were pooled and homogenized in a 50 mM TRIS-HCl buffer, pH 7.4, containing 150 mM KCl and 2 mM EDTA at a ratio of 1:4 (w:v). The homogenates were centrifuged at 9,000 *g* for 20 minutes. The total amount of protein in each supernatant was determined using the standard Coomassie protein assay (Bradford). Supernatant aliquots were kept à -80 °C until usage.

The incubations were performed as previously described [Lavoie *et al.*, 2013; Santamaria *et al.* 2014] and were performed minimally in triplicate. The incubations were performed in a microcentrifuge tube and contained various concentrations ranging from 1 to 100 μM of ketamine or midazolam, 0.5 mg/mL of S9 fraction proteins diluted in 100 mM phosphate buffer, pH 7.4. Liver S9 enzyme suspensions (total volume of 100 μL) were fortified with 5 μL of NADPH-regenerating solution A (Corning BD Biosciences Cat. No. 451220) and 1 μL of solution B (Corning BD Biosciences Cat. No. 451200) and preincubated at 37°C for 5 min prior the addition of  ketamine or midazolam. For the determination of $K_m$ and $V_{max}$, the concentration of norketamine or α-hydroxymidazolam was determined after 10 min incubation to calculate the initial rate of formation (i.e. $V_i$). Fifty μL of samples were taken and mixed with 250 μL of the deuterated internal standard solution (1 μM $d_4$-norketamine or $d_4$-α-hydroxymidazolam in acetonitrile) in a 1.5 mL centrifuge tube. Samples were centrifuged at 12,000 *g* for 10 min and 200 μL of the supernatant was transferred into an injection vial for HPLC-MS/SRM analysis.

***Quantitative Analytical Methods***

The concentrations of norketamine and α-hydroxymidazolam were determined using an HPLC-MS/SRM assay. Two μL samples were injected using a Thermo Scientific Accela HPLC System (San Jose, CA, USA) onto a Thermo Hypersil Phenyl 100 x 2 mm column (5μm) with flow rate of 300 μL/min. The mobile phase consisted of a mixture of acetonitrile, methanol, water and formic acid at a ratio of 60:20:20:0.1, respectively. The Thermo Scientific linear ion trap mass spectrometer (Thermo LTQ-XL) was interfaced with the HPLC system using a pneumatic assisted electrospray ion source. The mass spectrometer was coupled with the HPLC system using a pneumatically assisted electrospray ion source (ESI). The sheath gas was set to 25 units and the ESI electrode was set to 4000 V in positive mode. The capillary temperature was set at 300 °C, the capillary voltage to 15 V and the collision energy was set to 40%. All scan events were acquired with a 100 msec maximum injection time. Metabolites and corresponding deuterium-labeled molecule analogues were analyzed in full scan MS/MS and the quantification was based on specific post-processing SRM extracted ion chromatograms. The selected reaction monitoring (SRM) transitions were set to m/z 224.1 → 179.0, 228.1 → 183.0, 342.1 → 297.0 and 346.1 → 301.0 for norketamine, $d_4$-norketamine, α-hydroxymidazolam and $d_4$-α-hydroxymidazolam respectively. The analytical range used was ranging from 0.05µM to 50 µM.

***Data analysis and regression***

All non-linear regression analyses were performed with PRISM (6.0f) GraphPad software (La Jolla, CA) using the non-linear curve-fitting module with an estimation of the goodness of fit. The Michaelis-Menten equation describes the rates of irreversible enzymatic reactions that are generally observed for CYP mediated metabolic reactions. Michaelis-Menten parameters can be estimated with non-linear regression analysis using the Michaelis-Menten equation [Michaelis and Menten, 1913].

$$v_i = \frac{v_{max}\,[S]}{K_m + [S]} \quad (1)$$

Were the initial velocity ($v_i$) was determined using equation 2.

$$v_i = \frac{d[P]}{dt} = \frac{[\alpha\text{−hydroxymidazolam}]_{10\,min}}{10\,min} \; or \; \frac{[norketamine]_{10\,min}}{10\,min} \quad (2)$$

The initial rate ($v_i$) was calculated based on the concentration of norketamine or α-hydroxymidazolam measured after a 10 minutes incubation of ketamine or midazolam in rat liver S9 enzyme suspensions. Additionally, the enzyme-mediated clearance ($CLu_{int}$) that would occur without physiological limitations including protein binding or hepatic blood flow was determined using equation (5).

$$CLu_{int} = \frac{V_{max}}{K_m} \quad (5)$$

### *Statistical analysis*

Results are presented as means (± SD). The statistical difference was assessed with a one-way ANOVA and a Tukey's multiple comparisons test using GraphPad PRISM (version 6.0f); $p < 0.05$ was considered significant.

## Results and Discussion

In our previous study, an intraperitoneal administration of ketamine (80 mg/kg) and xylazine (10 mg/kg) in male Sprague Dawley rats of 3, 6, 12 and 18 months of age (n=6/age group) caused a significant increase of the $AUC_{0\text{-}t}$ and $AUC_{0\text{-}\infty}$ with aging [Giroux *et al.*, 2015]. Compared to 3 month old rats, the ketamine exposure ($AUC_{0\text{-}t}$) increased up to a factor of 4.15 compared with 18 month old rats. The elimination half-life ($T_{1/2}$) and relative clearance (CL/F) significantly decreased with aging. The decrease in $T_{1/2}$ and CL/F might be explained by an age-associated reduction in renal function or an increase in volume of distribution of lipid soluble drugs resulting in a prolongation of $T_{1/2}$ but we can't exclude hepatic clearance impairments. Ketamine is metabolized primarily by the liver CYP3A [Santamaria *et al.*, 2014; Mössner *et al*. 2011] (i.e. CYP3A1 and CYP3A2) to its active metabolite norketamine. Alterations in liver metabolism (e.g. hepatic clearance) with aging could explain the higher AUC observed in the aging rats. Moreover, selected organs, including liver, were taken for histopathology to confirm tissue integrity with aging.

The effect of substrate concentration on the initial rate ($V_i$) of an enzyme-catalyzed reaction is a fundamental concept in enzyme kinetics. $K_m$ is an indicator of the affinity that an enzyme has for a particular substrate, hence the thermodynamic stability of the enzyme-substrate complex. The stability of the enzyme-substrate complex is closely related to primary, secondary and tertiary structures of the enzyme. It plays a central role in defining the energetically favored binding cluster of the substrate into the active enzyme site [Sun and Scott, 2010]. As shown *in silico*, the structure of the binding cluster may lead to different metabolites or affect the rate of formation [Sun and Scott, 2010]. To adequately determine the value of the apparent Michaelis-Menten constant $K_m$ as well as the maximum rate achieved by the system ($V_{max}$), the data were fitted with the Michaelis-Menten equation (1). The initial rate ($V_i$) was calculated using the equation (2) and concentrations of α-hydroxymidazolam or norketamine were determined after 10 min midazolam or ketamine incubation in rat liver S9 fraction suspensions by HPLC-MS/SRM. Figure 2 shows results coherent with a kinetics following a Michaelis-Menten enzymatic reaction for all rat liver S9 fractions from all age groups and data were compatible with commercial rat liver S9 fractions (data not shown). The derived data are presented in Table 1. Midazolam is a well-characterized substrate of CYP3A and the primary biotransformation product is α-hydroxymidazolam. As illustrated in Table 1 and Figure 2, the observed $K_m$ values were not significantly different when comparing age groups. This is interesting because it suggest that enzyme-substrate complex structure was not significantly different between midazolam and CYP3A (i.e. CYP3A1 and CYP3A2) with age. However, the derived $V_{max}$ suggest a rapid saturation of the CYP3A enzyme active sites in liver S9 fractions of 18-month old rats, thus affecting significantly the intrinsic clearance ($CLu_{int}$) of the drug. Interestingly for ketamine, the derived $K_m$ value significantly changed in geriatric rats. This observation is very distinctive compared to rat liver CYP3A mediated metabolism of midazolam. The $K_m$ value is directly related to the thermodynamic stability of the binding cluster of ketamine in the active site of CYP3A. The data indicates a significant decrease of $K_m$ in liver S9 fractions of 18-month old rats and suggest that ketamine may have stronger interactions with CYP3A active site residues leading to a more thermodynamically stable enzyme-substrate complex. However it may also suggest that the enzyme binds the substrate more tightly and consequently necessitates more energy to form the activated transition state complex ($EX^{\ddagger}$), a necessary intermediate to form the enzyme-product complex (EP) as shown in Equation 6.

$$E + S \rightleftharpoons ES \overset{k_{cat}}{\rightleftharpoons} EX^{\ddagger} \rightleftharpoons EP \rightleftharpoons E + P \quad (6)$$

Free energy difference associated with the formation of the enzyme-substrate complex ($\Delta G_{binding}$) can have a significant impact on the observed $K_m$ but an increase of the $\Delta G^{\ddagger}$ (difference of free energy between $EX^{\ddagger}$ and ES) can also significantly decrease the rate constant $k_{cat}$ (i.e. turnover number) witch represents the number of substrate molecules each enzyme site can convert to a metabolite per unit of time. The rate constant $k_{cat}$ can be related to $V_{max}$ using the Equation 7.

$$V_{max} = k_{cat} \times [E_{total}] \quad (7)$$

As illustrated in Figure 2 and in Table 1, derived $V_{max}$ values shows significant differences with age specifically when comparing with results obtain from liver S9 fractions of 18-month old rats. A decrease of $V_{max}$ suggest a rapid saturation of the CYP3A enzyme active sites similar to what has been observed with the reference CYP3A substrate midazolam. These results are therefore compatible with the formation of a more stable enzyme-substrate complex (ES) but a less favorable transition state complex ($EX^{\ddagger}$) leading to an increase of $\Delta G^{\ddagger}$, thus a decrease of $k_{cat}$ and $V_{max}$. Conformational change of the CYP3A active site with age can potentially explain these results and protein misfolding is characteristic of several age-related problems. The interaction of the active site residues is substrate dependent and the results appear to suggest that an energetically favored binding cluster of ketamine in the active site of CYP3A is observed with age but interestingly this effect was not observed for midazolam. The formation of a more stable enzyme-substrate complex (ES) may have severe consequences on drug-drug interactions, a major issue in geriatric populations. Equation 5 was used to calculate the $CLu_{int}$, and the observed $CLu_{int}$ diminishes with aging, associated to the midazolam CYP3A catalyzed reaction, but increases for ketamine (Table 1). When using equation 5, the calculation of $CLu_{int}$ assumes that the concentration of the enzyme catalytic sites remain constant. This assumption cannot be made if conformational changes of the CYP3A active site occurs with aging. Consequently, the $CLu_{int}$ cannot be compared between age groups for the ketamine CYP3A mediated reaction. As shown in Figure 2, the rate of formation of norketamine undergoes a very rapid saturation in the liver S9 fractions of 18-month old rats suggesting that following a relatively high ketamine dose (80 mg/kg), drug metabolism is impaired leading to a significant increase of drug exposition (AUC) and elimination. These results are in accordance with our recent in vivo investigation [Giroux et al., 2015].

**Conclusion**

Age and related factors have a substantial effect on ketamine bioavailability that can be related to physiological changes but also to liver metabolism alterations. Our results suggest that rat CYP3A enzyme undergoes conformational changes with age particularly in our geriatric group (e.g. 18 month rats) leading to significant decrease in the rate of formation of norketamine. Moreover, results strongly suggest a severe impairment of CYP3A ketamine mediated metabolism.

## Acknowledgments

This project was funded by the National Sciences and Engineering Research Council of Canada (F.Beaudry discovery grant no. 386637-2010). The HPLC-MS/MS analyses were performed on instruments funded by the National Sciences and Engineering Research Council of Canada (F.Beaudry Research Tools and Instruments Grants no. 439748-2013). R. Santamaria received a scholarship from *Le Fond du Centenaire* (Faculté de médecine vétérinaire, Université de Montréal). The animal study was funded by CALAM/CALAS (P. Vachon 2014 research fund).

## Authorship Contributions

Each author's contributions to the manuscript are the following;

Participated in research design: Santamaria R, Vachon P, Beaudry F

Conducted experiments: Santamaria R, Giroux MC, Vachon P, Beaudry F

Performed data analysis: Santamaria R, Beaudry F

Wrote or contributed to the writing of the manuscript: Santamaria R, Vachon P, Beaudry F

All authors: No reported conflicts of interest. The research was funded by recognized Canadian research funding agencies**.**

**Table 1. Kinetic parameters associated with the formation α-hydroxymidazolam and norketamine in liver S9 fractions from aging rats.**

| **α-hydroxymidazolam** | $V_{max}$ nmol $min^{-1}$ $mg^{-1}$ | $K_m$ μM (nmol mL-1) | $CLu_{int}$ mL min-1 |
|---|---|---|---|
| 3 month liver S9 fractions | 0.247 (± 0.017) | 4.32 (± 0.26) | 0.057 |
| 6 month liver S9 fractions | 0.264 (± 0.021) | 6.22 (± 0.99) | 0.042 |
| 12 month liver S9 fractions | 0.216 (± 0.020) | 4.18 (± 0.05) | 0.052 |
| 18 month liver S9 fractions | 0.0398 (±0.003)[1] | 3.76 (± 1.45)[3] | 0.011 |
| **norketamine** | | | |
| 3 month liver S9 fractions | 2.39 (± 0.23) | 15.99 (± 4.28) | 0.150 |
| 6 month liver S9 fractions | 2.61 (± 0.18) | 18.38 (± 3.85) | 0.142 |
| 12 month liver S9 fractions | 2.07 (± 0.073) | 19.04 (± 0.94) | 0.109 |
| 18 month liver S9 fractions | 0.68 ( 0.02)[2] | 2.65 (± 0.49)[4] | *0.256* |

**[1,2,] $p < 0.001$**
**[3] $p > 0.05$**
**[4] $p < 0.05$**

**Figure legends**

**Figure 1.** Molecular structures of ketamine, norketamine, midazolam and α-hydroxymidazolam

**Figure 2.** Determination of Michaelis constant $K_m$ and maximum velocity $V_{max}$ using non-linear regression fitting. Each point represents the mean (±SD) of experiments in triplicate. Significant differences of the initial rate of formation ($V_i$) where observed starting at 5µM substrate concentration for liver S9 fractions of 18-month old rats.

*** p < 0.05**
**** p < 0.01**
****** p < 0.0001**

**Figure 1.**

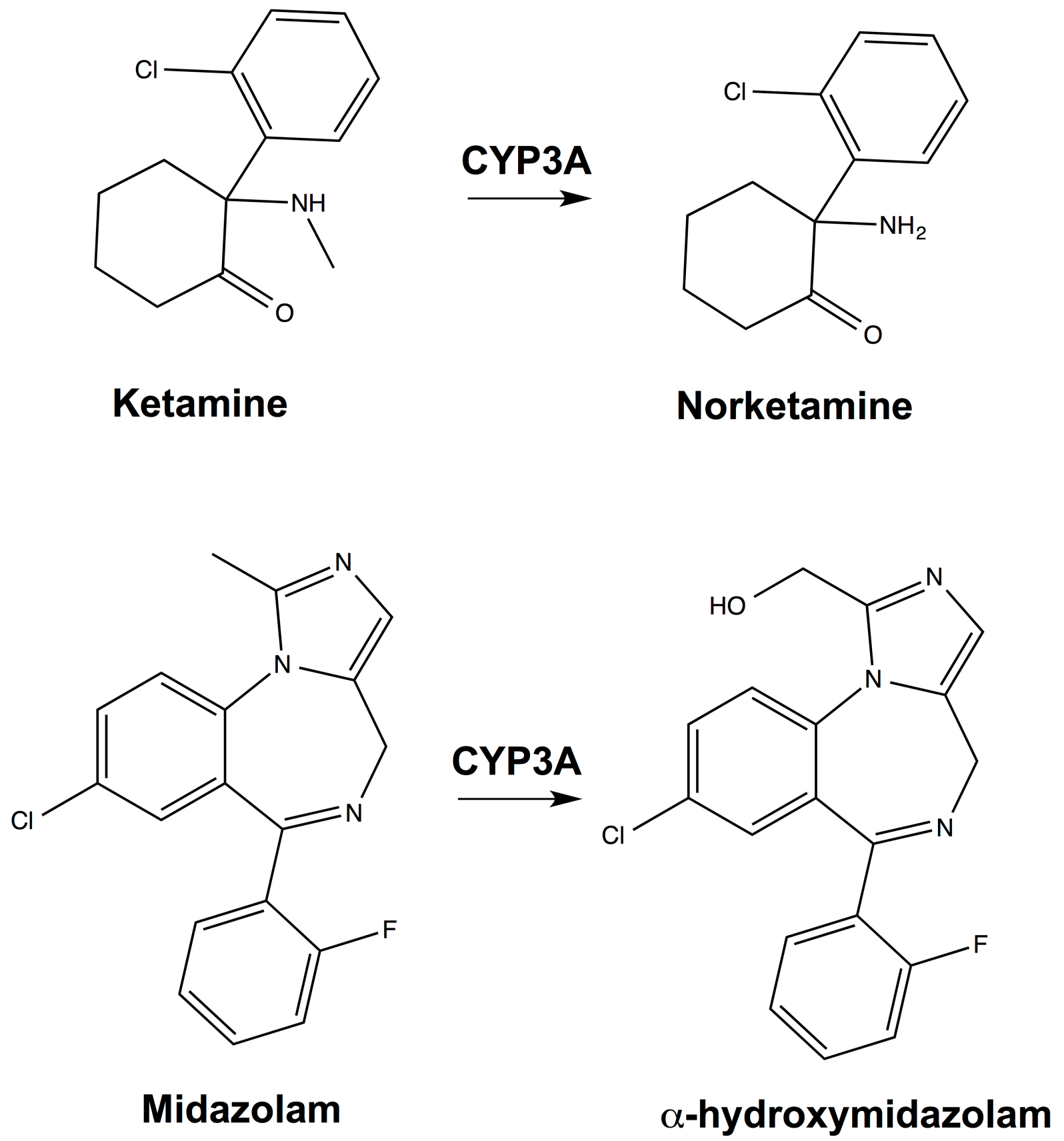

**Figure 2.**

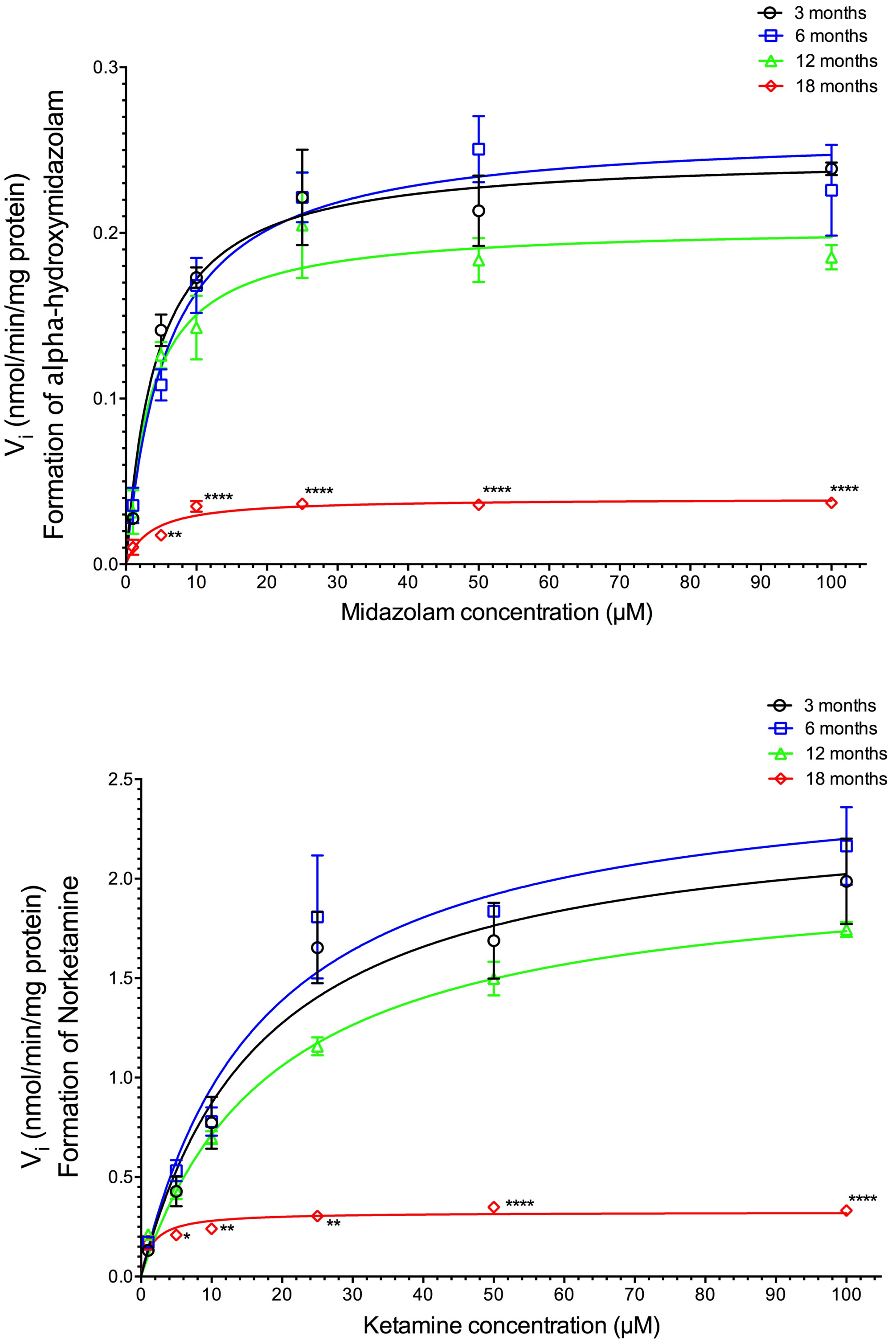